\documentclass[sigconf,10pt,nonacm]{acmart}
\usepackage{caption}
\usepackage{subcaption}

\title{20 Years in Life of a Smart Building: A retrospective}

\author{Karolina Skrivankova}
\affiliation{%
  \institution{UCL}%
  \city{London}
  \country{UK}%
}
\author{Mark Handley}
\affiliation{%
  \institution{UCL}%
  \city{London}
  \country{UK}%
} 

\author{Stephen Hailes}
\affiliation{%
  \institution{UCL}%
  \city{London}
  \country{UK}%
} 

\begin{abstract}
Operating an intelligent smart building automation system in 2025 is met with many challenges: hardware failures, vendor obsolescence, evolving security threats and more. None of these have been comprehensibly addressed by the industrial building nor home automation industries, limiting feasibility of operating large, truly smart automation deployments. This paper introduces KaOS, a distributed control platform for constructing robust and evolvable smart building automation systems using affordable, off-the-shelf IoT hardware. Supporting control applications and distributed system operations by leveraging containerisation and managed resource access, KaOS seeks to achieve flexibility, security, and fault tolerance without sacrificing cost-effectiveness. Initial evaluation confirms the practical feasibility of our approach, highlighting its potential to sustainably maintain and incrementally evolve building control functionalities over extended timeframes.

\end{abstract}

\begin{document}

\maketitle

\section{A retrospective}

Way back in 2026 we moved into a brand new ``smart'' building.  
To cut costs, the university chose to use cheap off-the-shelf wireless IoT
hardware throughout.  We called it {\em kaosnet} as it ran on a new
platform called KaOS.  Many were skeptical at the time given how
IoT was back then, but looking back, 2026 was a turning point. Kaosnet
has weathered a great deal of change and still works well today.  But
then there's not much of the original kaosnet left.

First to fail were the motion sensors. We had redundant
sensors everywhere - they were really cheap - so initially we didn't
notice. Kaosnet just reconfigured to use the working ones.  We swapped
the failed ones out when we found time, but then the vendor went
bankrupt, so we had to replace them with different ones.  By then
several companies supported KaOS so we ran the same control apps on them.

In 2029 the company next door moved out and we integrated their
building with ours.  Of course none of the original hardware was
available, so we now had a completely heterogeneous network.
With great foresight, the university decided to redo the floor layout
just 6 months later, leaving gaps in the motion detection.
We added new passive radar sensors - they integrated straight in and ran
the same apps.

The great fluxnet worm of 2034 was a big problem for many.  Kaosnet
was never cloud-based but some of our lights became
infected from a student's headset via their LiFi interface.  Fortunately
kaosnet confined the problem to the kaos LiFi app, and no other
systems were infected. The lighting vendor no longer issued security
patches, but the same app was used by many vendors and still supported,
so no big deal.

We deployed our first robots in 2037.  They were great at many
things but opening doors was not one of them.  We integrated powered
doors, but the robots triggered security alarms, so we swapped some of
the motion sensors for cheap AI cameras.  They supported kaos and ran the
same app logic to turn the lights on, but now only for humans.  We
added the robots to kaosnet too, so they didn't trigger the heating at
night as they cleaned; fortunately they run kaos apps like everything
else.

2041 wasn't a great year - we had a fire one night in the basement
network hub.  The robots put it out quickly, but the wireless APs were
left disconnected from each other.  Kaosnet's decentralized control
loops carried on working until insurance eventually sorted out the
mess.

We do still have some 20-year old thermostats on kaosnet, but
that's all, and we replaced their software for the 2038 Unix clock
rollover.  The network has many more devices now than it did
back then, from hundreds of vendors.  And it all still works. This
paper from 2025 describes how.

\section{State of industrial and home automation in 2025}
The industrial and home building automation industries share a common goal: transform passive buildings into smart environments that increase user comfort and use resources efficiently. Neither industry has successfully enabled building automation to reach its full potential; encountering distinct challenges. Industrial systems are robust but rigid, whereas home automation systems are more flexible but are insecure and unreliable. We argue that for such systems to remain effective on decade timescales we need the best of both worlds. 

Industrial building automation deployments have traditionally prioritized durability, reliability and stable control.  They rely predominantly on robust hardware and wired connections designed for longevity\cite{Artetxe_Barambones_Calvo_Fernández-Bustamante_Martin_Uralde_2023}. Interoperability is not a problem, just so long as solutions are based on a single vendor's proprietary hardware and software.  However, this very reliability leads to rigidity and obsolescence; the high initial investment and the complexity of deployment create networks that are resistant to change. Vendor lock-in, high incremental costs and a cautious approach to innovation tend to result in rudimentary control policies.  Upgrading these systems or integrating new technologies typically involves high costs, specialized expertise, and significant downtime.  Such rigidity makes it hard to progressively implement and incrementally improve the sort of tailored, sophisticated control policies that are needed to improve energy efficiency and user comfort as building use changes over time\cite{Brooks_Coole_Haskell-Dowland_Griffiths_Lockhart_2017}. 

Home IoT solutions target similar objectives but greatly reduce installation costs through cheap hardware and wireless networking. These cost advantages come at the expense of reliability and security\cite{Xing_2020}. Moreover, the consumer IoT ecosystems are built on individual devices and their respective connectivity,  rather than integrated control systems.  This limits interoperability and makes it hard to support complex, multi-device and multi-agent control policies\cite{Taboada-Orozco2024}. The resulting fragmented ecosystem imposes a significant maintenance burden, requiring manual management of device failures, software updates, and connectivity disruptions\cite{Sivasankari_Rathika_2025}.

Future building control systems must accommodate higher densities, more granular and adaptive control, and dynamic integration of new technologies. As the complexity and scale increases, systems need to evolve flexibly, where the software and hardware components are replaceable and upgradeable independently throughout the lifetime of the network.  This allows functional requirements to change over time, and component failures and obsolescence to be managed\cite{Ożadowicz_2024}. The increasingly dynamic evolution of all these components also introduces new security challenges, and requires robust strategies to defend against vulnerabilities.

Centralized automation systems, like Home Assistant\cite{home-assistant}, solve some of these problems by putting all the intelligence in one place.  A single controller manages automation logic, often using YAML configurations and MQTT, and a diverse open ecosystem of plug-ins eases deployment. While convenient initially, this design inherently introduces critical single points of failure, limiting resilience and adaptability\cite{wang_understanding_2022}. Even efforts such as the Matter standard\cite{matter} and Thread protocol\cite{openthread}, which significantly improve device compatibility through common IP-based communication, do not fundamentally address decentralized fault tolerance or system evolvability. These centralized architectures remain complex to maintain and coordinate over extended lifetimes.

We argue that it is possible to leverage the improved capabilities of cheap unreliable embedded hardware platforms to build a distributed control system that can \textbf{adapt and evolve over decades} while providing a \textbf{reliable} and \textbf{secure}\textbf{ infrastructure} for control applications. 
Our approach builds upon but diverges from prior automation solutions. 

Recent lightweight virtualization techniques, such as FemtoContainers \cite{Zandberg_Baccelli_Yuan_Besson_Talpin_2022} and Web Assembly\cite{Haas_Rossberg_Schuff_Titzer_Holman_Gohman_Wagner_Zakai_Bastien_2017}, demonstrate the potential for executing portable, sandboxed control logic directly on existing embedded devices as cheap as an ESP32\cite{esp32}, highlighting the feasibility of distributed approaches. Our architecture uniquely combines this containerized runtime model, leveraging containerised control tasks for efficient, platform-agnostic deployment, with a decentralized, local-first execution strategy. This design explicitly targets long-term evolvability, robustness, and cost-efficient incremental upgrades, overcoming the inherent limitations found in both centralized automation systems and traditional BMS.

\section{Life of an automation system}

Consider a future building automation system that adapts to the needs of users as they change over both the short and long term, seeking to maximise operational efficiency and minimize resource consumption and environmental impact. It uses a fine-grained spatial network of mostly low-cost off-the-shelf devices with sensing and actuating peripherals placed throughout the building. There is no central controller---the devices themselves host much of the logic making up the control architecture needed to achieve high level goals. The system was commissioned in 2025 and has been in operation for a decade. 

The control architecture of the building and the control systems that it supports have had to change frequently since their initial commissioning: the underlying physical layout was adapted to changing occupation density and desired functionality, such as transforming a set of offices to a workshop with a controlled environment. Whereas, in the past, these changes would happen perhaps once a decade, modular architectural design and the tight integration of robotics into the environment have enabled the use of space to adapt as required on both micro and macro levels. Changes in daily building use patterns, operational requirements, device failures and hardware availability mean that we need to adjust both the desired behaviour and the underlying implementation over time.


\subsection{Device evolution}
The continuous correct execution and adjustment of building control systems is challenging when trying to maintain a decentralised system made up of thousands of heterogenous devices. To accommodate change, we need the system to be capable of evolution on both the device and system (network of devices) levels. 

Consider a device that runs two tasks: \textit{collect  data} and \textit{calculate average} (Figure 1). After 10 years of operation, this device dies and the wider system loses both the data stream that was supplied by the device and the compute used to carry out the two control tasks. With device failures becoming common as the system scales, an important question is how can we effectively deal with such losses?

\begin{figure}[!b]
    \centering
    \includegraphics[width=0.68\linewidth, trim={0 2mm 0 1mm},clip]{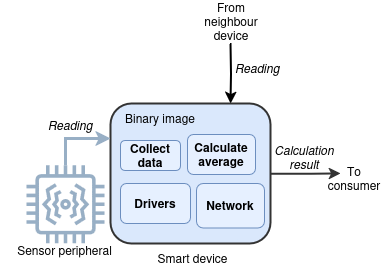}
    \caption{\vspace{-2mm}The failed device}
    \label{fig:enter-label}
\end{figure}

Feedback loops in control systems contain both explicit and implicit dependencies between components.  Due to these dependencies, faults in components of a control architecture sometimes cause cascading failures, increasing disruption. Mitigation strategies important for fault tolerance, such as redundancy, are only possible if the additional cost of extra equipment can be justified.  Often the hardware and installation costs are too high (as in industrial building automation), or the integration and maintenance too painful (home automation).  

In Figure 1, the data collection task collects sensor readings and sends them to the average computation task, which fuses local data with that from a neighbouring sensor as part of a larger control loop. Sensing is fundamentally bound to the device but averaging is not---it could be performed anywhere---but running it on the sensor device minimizes external dependencies, reducing latency and improving resilience.  
 But, when the device fails, the new replacement device is unlikely to run the same firmware, or even to come from the same vendor---especially so considering the short product lifecycles typical for IoT devices.  We need to ensure we can integrate replacement devices easily and securely. 

To maintain system capabilities over time, we would like to be able to run the existing `old' software such as the collection and averaging tasks on the `new' host device (Figure 2(a)), rather than having to run the new vendor's utility firmware.  At the same time, when a security problem comes to light with the old averaging software, we want to be able to upgrade it, which requires that the software is used and supported by someone other than the now-defunct hardware vendor.  Further, to allow capabilities to evolve and improve over time, an old device needs to be capable of running different new software, such as replacing the Average task with a Median task to improve outlier rejection (Figure 2(b)).

\begin{figure}[!h]
    \centering
    \begin{subfigure}[b]{0.3\textwidth}
        \centering 
        \includegraphics[width=\textwidth]{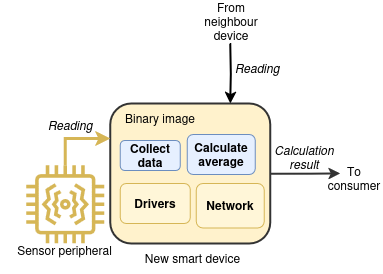}
        \caption{New device, old software}
        \label{fig:enter-label}
    \end{subfigure}

    \begin{subfigure}[b]{0.3\textwidth}
        \centering
        \includegraphics[width=\textwidth]{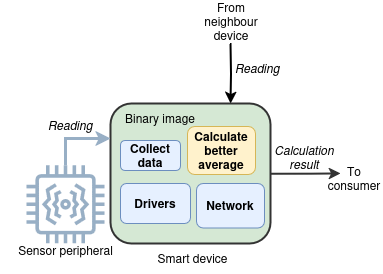}
        \caption{Old device, new software}
        \label{fig:enter-label}
    \end{subfigure}
    \caption{Device software}
\end{figure}

Tight coupling of software and hardware ecosystems, especially coupling the control architecture to vendor firmware, causes network operators to rely on a restricted number of vendors to minimize operational costs.  This, in turn, strengthens vendor lock-in and ossification of the ecosystem, with little incentive to innovate.  In such a world, customers have no choice but to stay or pay to refurbish the entire network. Development of new control software without decoupling requires hardware vendor support, or access to maintained device-specific source code; neither is likely on longer timescales. 

Ideally, especially when assigning multiple control tasks to a single device, we would like the ability to change software tasks dynamically at runtime.  For example, when the motion sensor device fails, the averaging task could be moved to the neighbouring device, minimizing disruption to the larger control loops these devices form a part of. In contrast, in 2025 smart devices usually run full-stack locked-down vendor software, with little room for the user/operator to adjust either device behaviour or its role within the larger control system. 

Thus, for both the device-software interactions shown in Figure 2, we benefit from decoupling software from hardware, especially when it comes to the functions comprising the higher-level system control architecture (collect data, calculate average). In an ideal world, we have multiple parallel independent ecosystems that allow for the update of old devices to new behaviour and vice versa.   Putting all this together, we would like device software to evolve incrementally as required, as shown in Figure 3(b) rather than the current status quo in Figure 3(a).  Our approach, then, is effectively a distributed operating system that supports the deployment of containerised control tasks on cheap low-powered devices, providing a common control layer interface to support the flows of data between tasks that comprise all the higher-level control loops.

\begin{figure}[!h]
    \centering
    \begin{subfigure}[b]{0.4\textwidth}
        \centering
        \includegraphics[width=\linewidth]{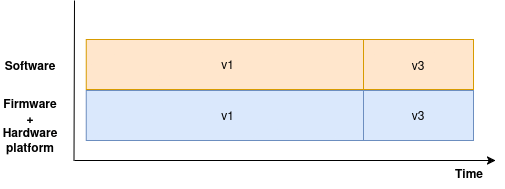}
        \caption{Software and hardware coupled}
        \label{fig:software-coupled}
    \end{subfigure}
    
    
    \begin{subfigure}[b]{0.45\textwidth}
        \centering
        \includegraphics[width=\linewidth]{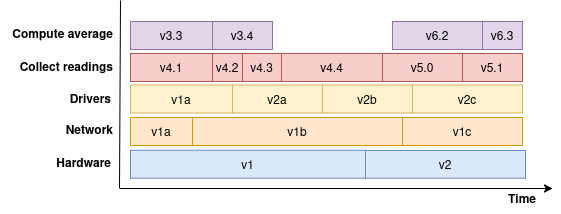}
        \caption{Evolvable ecosystem}
        \label{fig:sfotware-uncoupled}
    \end{subfigure}

    \caption{Software evolvability}
\end{figure}

\subsection{Network of devices}
While building automation networks are built of devices, it is only when they come together in a network implementing distributed control loops that useful work can be done (Figure 4(a)).  An individual device may need to to adapt its own behaviour depending on the state of other devices in the system, for instance when a new source of data is to be added to average calculation for improvement of inference accuracy. Or when a device becomes inaccessible, due to a wireless link interruption, hardware failure, or software fault. Software evolvability allows a device to flexibly adjust its functionality during its lifetime. But to maintain a stable control system, it is the network of devices making up the control system that has to evolve.

\begin{figure}[!h]
    \centering
    \begin{subfigure}[b]{0.4\textwidth}
        \centering
        \includegraphics[width=\linewidth]{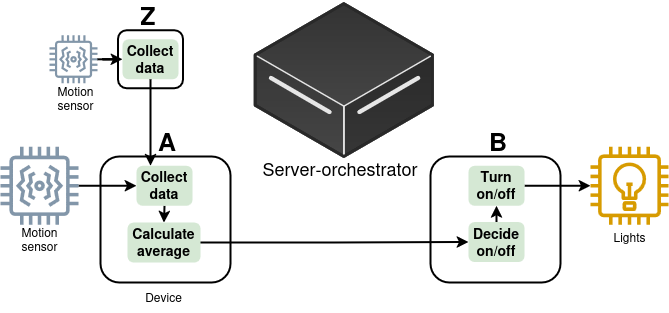}
        \caption{Network example}
        \label{fig:enter-label}
    \end{subfigure}

    \begin{subfigure}[b]{0.4\textwidth}
        \centering
        \includegraphics[width=\linewidth]{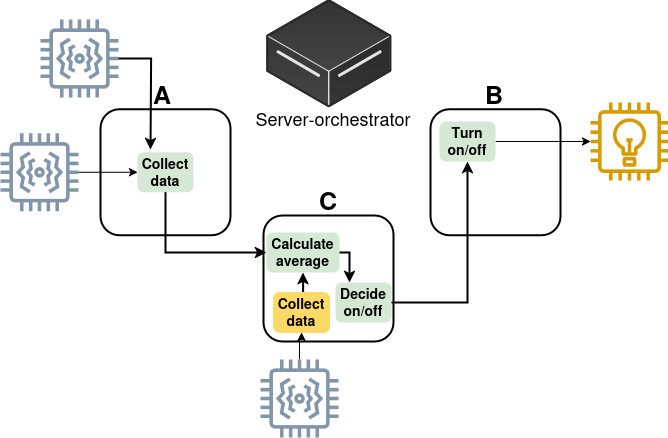}
        \caption{Add a device}
        \label{fig:enter-label}
    \end{subfigure}
    
    \begin{subfigure}[b]{0.4\textwidth}
        \centering
        \includegraphics[width=\linewidth]{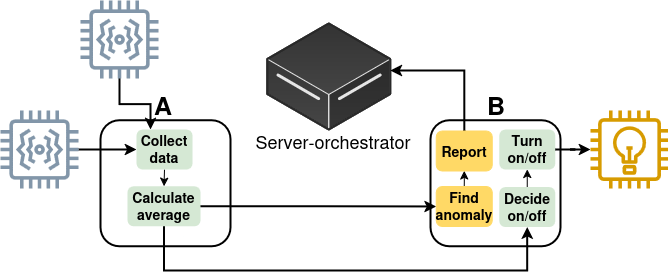}
        \caption{Extend functionality}
        \label{fig:enter-label}
    \end{subfigure}
\caption{Network of devices}
\end{figure}

To adapt system behaviour, we need to consider how the network of devices participates in control architecture execution. Devices will be added or removed from the network, intentionally or due to faults. When that happens, we may want to:
 \begin{enumerate}
    \item Add new functionality and improve performance as devices are added
     \item Retain old functionality as devices are lost  
     \item Degrade functionality gracefully as devices are lost 
     \item Extend functionality using the existing set of devices
 \end{enumerate}


\subsection{Add new functionality and improve performance as devices are added}
Consider adding a new device C to the example in Figure 4(a). This new device hardware platform is more powerful, with more accurate proximity sensors. Adding this device could improve the ability of the network to control the lighting based on occupancy.  What do we need to properly integrate it into our existing control architecture?

We can think of the control architecture as a directed graph of logical blocks of compute connected by data flows (top of Figure 5). The blocks delineate processing steps and the handover of data in the control architecture. These are mapped (ignore how for now) onto the currently available set of devices to bring about the desired behaviour of the system. When adding a new device node in our example, we need to add a new logical block node 'collect data' which must run on new device C because it has the sensor peripheral.  Then we need to establish a new communication channel to the `calculate average` task, which processes information for the tasks further down the pipeline. These operations are prompted via a server/orchestrator (could be multiple, representing different vendors responsible for different applications in the system) not involved in the critical control loops, but which facilitates privilege management to limit disruptive potential if a device hosts malicious software.    

The task `collect data' must be assigned to a device with the required hardware, in this case a motion sensor, but this is not the case for all tasks. For instance, `calculate average` could be placed on any of the devices, but not all assignments may be equally desirable. Link quality might make device C a better candidate for processing readings from A and Z, and its more capable hardware might make placing more compute on the device more sense than on the older device A. Importantly, these conditions change over time.  

To adapt the control architecture, we can manipulate its graph by adding or removing nodes and links, and adjust its mapping onto the hardware plane to reflect current conditions and requirements. 
The smaller the set of tasks with specific hardware dependencies, the more flexibility the system has to adapt the mapping. For instance, redundancy can be established by initialising a new node and either mapping it to a new device, as in Figure 4(b), or to an old device where copies are not co-resident. On the other hand, so long as we can maintain some minimal live graph with a minimal number of compute nodes live and reachable, then the system can execute the desired behaviour.

Building a control system from low-cost devices de-risks the purchase of extra equipment, and software evolvability decreases the maintenance burden posed by introducing extra hardware into the system. Together, they transform extra equipment from a liability into an asset.  They simplify scaling of the system and add extra capacity to deal with failure. By providing more options to host such minimal graph, they reinforce the system's ability to tolerate faults by creating extra capacity for flexible control policy operation.

\begin{figure}
    \centering
    \includegraphics[width=0.5\linewidth]{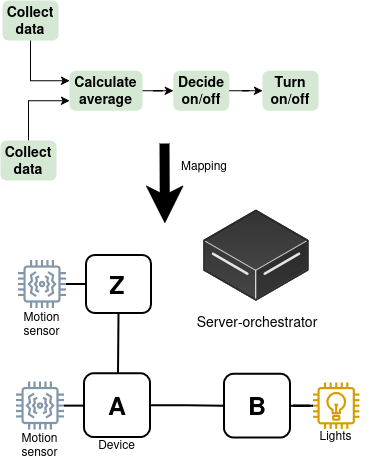}
    \caption{Find a mapping from control task set (directed graph) to devices connected over wireless network links}
    \label{fig:enter-label}
\end{figure}

\subsection{Retain functionality as devices are lost}
If device C is lost, we lose a crucial piece of executable logic: the `calculate average' task. Without it, collected readings cannot be processed and handed to the decision task. Luckily, this can be remedied, as this task does not have hardware peripheral dependencies, so in theory, we should be able to revert the control architecture graph from that in Figure 4(b) back to the original state in Figure 4(a).  

This brings us back to software evolvability. To enable this, we need the ability to execute control tasks on any of the many hardware platforms present in our system (hardware portability), and to securely load and execute it on the target device (dynamic code portability). Containerisation provides us with the former and leads us towards the latter. 
 
In our approach, control tasks are encapsulated within WebAssembly containers---a hardware-agnostic compilation target. This allows us to isolate tasks from one another and from the host firmware, providing a lightweight, secure sandboxed execution environment~\cite{Haas_Rossberg_Schuff_Titzer_Holman_Gohman_Wagner_Zakai_Bastien_2017}. Sandboxing provides a security boundary, improving system security by limiting the potential damage from vulnerabilities or malicious code.  This is crucial because dynamic code portability as discussed here opens a new attack surface. Our platform, KaOS, reinforces this with guarded capability-based container lifetime management. We chose WebAssembly for KaOS due to its portability and wide hardware platform support, but the approach is runtime-agnostic. Alternative container runtimes such as FemtoContainers\cite{Zandberg_Baccelli_Yuan_Besson_Talpin_2022} could be used instead.\footnote{Whereas WebAssembly does support many embedded platforms, many of these are mostly on the more capable side.}

By isolating control tasks within containers, third-party developers can safely contribute optimization algorithms without needing full device access, thus fostering an open ecosystem. Access to source code is not explicitly required, so there are incentives for innovation and development. 
Containerisation allows old functionality to be retained by re-assigning lost tasks to available hardware, thus maintaining the minimal graph of compute necessary for behaviour execution.  This is a core building block of KaOS.

\subsection{Degrade functionality gracefully}
The loss of some devices and peripherals can be tolerated via re-assignment of control tasks, but at a certain point the system loses the ability to cope due to fundamental limitations of available hardware: loss of sensing or actuating capacity (in the example, insufficient motion sensors to maintain an accurate notion of room occupancy, or loss of heaters leaves insufficient heating capacity to bring temperature up to a set-point). At this point, the desired response from the system is to gradually reduce service where possible. For instance, loss of motion sensing may mean the lighting cannot be perfectly controlled based on occupancy, but it can fall back to other mechanisms such as using $CO_2$ readings to determine people are still present despite gaps in motion coverage. 


As described above, we use containerisation as a key ingredient needed for such a flexible control architecture management. However, some containers need to communicate with other control containers to supply or consume data as part of feedback loops, locally or over the network.  Others require access to external resources such as hardware peripherals.  
WebAssembly does not provide a well-defined system call interface, but we need a simple future-proof interface which limits the ability of a malicious control container to affect the system outside of its runtime.  As containers are re-assigned when system conditions change, the dataflow source and destination are also expected to change (Figure 4(a) to Figure 4(b): dataflow between `collect' and `average' blocks). Building control systems are, by their nature, prone to cascading failures due to dependencies between control blocks. This makes it crucial for system stability, so that we can isolate failure, preventing it from spreading to neighbours.

To this end, KaOS uses abstract message-passing communication channels to transfer data between containers, whether locally or over the network. A container does not have the ability to initialise a communication channel by itself, limiting the effects of bugs or exploits.  Instead, containers are started by the KaOS orchestrator with a pre-approved set of communication channels.  However, this means the message channel needs to fulfil service needs for common control applications. Here, the peculiar nature of the building control system comes to our benefit: 
\begin{enumerate}
    \item lenient real-time characteristics: applications rarely require response time below 100ms, usually 
    \item distinct data categories with related delivery guarantee requirements: periodic and event-based transmission for sensor data, irregular command and event data
\end{enumerate}

Hence, instead of direct network access, we provide communication channels with a set of service guarantee interfaces that containers can access for inter-control-task communication, supporting topology changes of the control task graph and its device mapping. 
KaOS natively supports distributed operations—such as reliable container migration from device A to C (Figure 4)—to prevent race conditions and reduce service disruption. These communication channels are managed outside the container runtime, serving as a protective boundary against in-container faults and buffering any unprocessed or unsent messages.

KaOS communication channels can be reliable or unreliable.  If reliable, it can be a \textit{push-channel} (similar to pipes), or it can be a \textit{pull-channel}.  The latter is useful to minimize latency across a chain of containers, when the tail of the chain needs to control the clocking of data from the head without building queues all along the chain, or for power-saving nodes only intermittently available for data transmission.

\subsection{Extend functionality with existing set of devices} 
In Figure 4(c) an anomaly detection module is inserted into an existing system.  With the KaOS platform, such extensions are typically handled by the addition of new control task nodes and setting up appropriate communication channels. 

The KaOS architecture provides an abstraction layer to enable \textit{location-independent} communication channels between distributed control tasks, explicitly supporting the dynamic graph structure and container lifecycle management described previously.  Specifically, this abstraction addresses the limitations of WebAssembly by providing a minimal but sufficient interface for communication between containers, access to hardware resources, and interactions with underlying platform functions.

This approach allows us to make use of lifetime management and failure mitigation strategies, such as progressive roll-out of updates, and redundant task copies, as needed. By responding flexibly to conditions, KaOS can prevent undesirable gradual system performance degradation, when operational debt causes an inability to fulfil control architecture requirements, even with available hardware capacity.   
Containerization also simplifies dynamic software updates, giving KaOS the capability to securely add, stop, or replace control task containers at runtime.

\section{Conclusion}
We argue that to realize genuinely smart, resilient, and adaptable automation systems, a shift in IoT software architecture is needed. KaOS exemplifies this by emphasizing evolvability, security, and fault tolerance, key interconnected properties necessary for long-lived, affordable, and reliable systems. Preliminary tests of running a simple control architecture on multiple cooperating ESP-32 devices have validated the feasibility of the containerization approach on small cheap devices.  We measure negligible execution latency added due to container crossing and processing overhead. Future evaluation will validate KaOS’s scalability, robustness, and suitability for sustained operation in an emulated environment of network sizes in the 100s - 1000s of devices, including systematic device integration, fault injection and cybersecurity scenarios.

\bibliographystyle{ACM-Reference-Format} 
\bibliography{reference}

\end{document}